\documentclass[a4paper,12pt]{article}
\pdfoutput=1

\usepackage{jheppub}
\usepackage[utf8]{inputenc}
\usepackage{amsmath,amssymb,amsthm}
\usepackage{physics,mathtools}
\usepackage{tikz,float}
\usepackage{bbm}
\usepackage{subcaption}
\usepackage{booktabs}

\usepackage{amssymb}
\usepackage{pifont}
\usepackage{tikz}
\usetikzlibrary{arrows.meta,positioning}

\makeatletter
\def\@fpheader{\vspace{0.1mm}}
\makeatother
\newcommand{\bw}{\begin{widetext}}
\newcommand{\ew}{\end{widetext}}
\newcommand{\bea}{\begin{eqnarray}}
\newcommand{\eea}{\end{eqnarray}}
\newcommand{\be}{\begin{equation}}
\newcommand{\ee}{\end{equation}}
\newcommand{\bca}{\begin{cases}}
\newcommand{\eca}{\end{cases}}

\newcommand{\ben}{\begin{enumerate}}
\newcommand{\een}{\end{enumerate}}

\usetikzlibrary{patterns}
\makeatletter
\let\over\@@over
\makeatother

\title{{{Holographic Correlators from Thermal Bootstrap}}}
\author{Ilija Buri\'c,}
\emailAdd{burici@tcd.ie}
\author{Ivan Gusev,}
\emailAdd{gusevi@tcd.ie}
\author{Andrei Parnachev}
\emailAdd{parnachev@maths.tcd.ie}

\affiliation{School of Mathematics and Hamilton Mathematics Institute, Trinity College, Dublin 2, Ireland}

\abstract{
Holographic thermal two-point functions can be analyzed using the operator product expansion  which contains contributions from both multi-stress-tensor and double-trace operators. The former can be computed by analyzing the bulk equation of motion in a near-boundary expansion, but the latter has remained elusive—in practice, one resorts to solving a partial differential equation with limited accuracy. We show that imposing the Euclidean periodicity condition on the holographic correlator (also known as the KMS condition or thermal bootstrap), followed by Padé–Borel resummation, provides an efficient method for computing double-trace thermal coefficients. The resulting series converges rapidly and yields numerical values in excellent agreement with those obtained from solving the partial differential equation.
}

\begin{document}

\maketitle

\section{Introduction and Summary}

The conformal bootstrap, \cite{Ferrara:1973yt,Polyakov:1974gs,Mack:1976pa}, is a program aimed at determining the set of allowed conformal data based on the unitarity of the theory and the crossing symmetry of CFT correlators. It has recently seen renewed interest and has led to improved understanding of CFTs, \cite{Rattazzi:2008pe,Poland:2010wg,El-Showk:2012cjh,Fitzpatrick:2012yx,Komargodski:2012ek,El-Showk:2014dwa,Simmons-Duffin:2016gjk,Caron-Huot:2017vep,Poland:2018epd,Chester:2019ifh,Rychkov:2023wsd,Chang:2024whx}.

A version of the conformal bootstrap is applicable to CFTs at finite temperature $T = \beta^{-1}$ and is based on the periodicity of thermal correlators in Euclidean time, $\tau \sim \tau + \beta$, \cite{El-Showk:2011yvt,Iliesiu:2018fao,Iliesiu:2018zlz,Alday:2020eua,Marchetto:2023xap,Barrat:2024aoa,Barrat:2025wbi,Buric:2025anb,Barrat:2025nvu} (see also \cite{Petkou:2018ynm,David:2023uya,David:2024naf,David:2024pir,Kumar:2025txh} for related works on the thermal inversion formula). Consider the two-point function of scalar operators. The leading short-distance singularity is governed by the contribution of the identity operator to the operator product expansion (OPE). Due to the Euclidean time periodicity, 
thermal images of this singularity reflect the contributions of operators with large conformal dimensions.\footnote{Note that the leading large-dimension behavior of thermal coefficients (which are essentially products of OPE coefficients and thermal one-point functions) is not necessarily governed by the KMS poles at $\tau = \pm \beta$. For example, in holography the leading behavior comes instead from the bouncing singularity \cite{Ceplak:2024bja} and may exhibit oscillating signs. However, such behavior is expected to cancel in the averaged one-point functions. \cite{Marchetto:2023xap}.}

In the simple example of generalized free fields (GFF), finite temperature correlators are obtained by summing the zero-temperature result over thermal images. The additional terms give rise to non-vanishing contributions from double-trace operators. Put differently, double-trace operators acquire expectation values that are precisely tuned to ensure that the full correlator is periodic in Euclidean time. The GFF result is analytic and free of singularities in the complex strip $0 < \mathrm{Re} \ \tau < \beta$.

One may wonder whether a similar picture holds for holographic correlators. The OPE for such correlators includes the identity, the stress tensor and its composites (the stress-tensor sector), as well as double-trace operators. The stress-tensor sector contribution can be computed by analyzing the bulk scalar equation of motion near the AdS boundary, \cite{Fitzpatrick:2019zqz}\footnote{See \cite{Karlsson:2019qfi,Li:2019tpf,Kulaxizi:2019tkd,Fitzpatrick:2019efk,Karlsson:2019dbd,Li:2019zba,Karlsson:2020ghx,Li:2020dqm,Fitzpatrick:2020yjb,Grinberg:2020fdj,Karlsson:2022osn,Parisini:2022wkb,David:2022nfn,Huang:2024wbq,Haehl:2025ehf} for computations of thermal coefficients in other holographic examples.
}. In contrast, computing the double-trace contribution is technically much more challenging. These terms appear as contact (delta-function–like) contributions in momentum space, and determining them requires numerically solving a partial differential equation in the bulk. This has been attempted in \cite{Parisini:2023nbd}, where the lowest-lying thermal double-trace coefficient was computed, but extending this calculation further remains technically difficult.

Can a simple summation over thermal images provide an algebraic method for determining the conformal data of double-trace operators in the holographic case? In \cite{Buric:2025anb}, we showed that getting the correlator which would be free of singularities in the complex strip is more intricate. Borel resummation of the formal solution to the Euclidean periodicity (KMS) condition provides the correct solution \cite{Buric:2025anb}. In the same work, we also computed thermal coefficients for double-trace operators in an asymptotic model, where the multi-stress-tensor data is specified by its large-dimension asymptotics.

In the present paper, we show that the method developed in \cite{Buric:2025anb} can be employed to efficiently produce accurate results for the double-trace thermal data in the full holographic correlator.\footnote{Note that we do not expect the thermal bootstrap or KMS condition to fix all thermal coefficients. For instance, in addition to Einstein–Hilbert gravity, one may consider higher-derivative gravitational terms in the bulk, which would lead to a different set of multi-stress-tensor thermal coefficients. Although the operator spectrum remains the same, the double-trace thermal coefficients adjust to different values in each theory to satisfy KMS periodicity.} We employ Padé–Borel (PB) resummation, together with the thermal data for multi-stress tensors, to compute the double-trace thermal coefficients. The convergence is rapid. Moreover, we improve on the approach of \cite{Parisini:2023nbd} by numerically solving the relevant partial differential equation to obtain independent values for the low-lying double-trace thermal coefficients. The results show excellent agreement, better than 1$\%$ precision, within the regime where the numerical PDE solution is reliable.

To summarize, the main result of this paper is a computationally efficient method for obtaining the thermal coefficients of double-trace operators in holographic CFTs. Along the way, we improved the numerical solution of the PDE governing the bulk scalar equation of motion; however, we regard this approach  as a consistency check, since the Pad\'e-Borel algebraic technique is significantly more efficient and precise. Finally, we also made progress in refining the procedure for determining the multi-stress-tensor thermal coefficients, which enabled us to compute them much faster.

The rest of the paper is organized as follows. In Section \ref{SS:stress-tensor-recursion}, we review the procedure of \cite{Fitzpatrick:2019zqz} for computing the thermal coefficients of the stress-tensor sector. We also recast it as a set of recursion relations. After reviewing the solution to the KMS sum rules  \cite{Buric:2025anb} in Section \ref{SS:Double-trace coefficients from the KMS condition}, we develop the Pad\'e-Borel method and apply it to find holographic double-trace coefficients in Section \ref{SS:Padé-Borel}. In Section \ref{S:Matching to the PDE solution}, these coefficients are compared to ones obtained from the numerical solution of field equations in the bulk. Section \ref{S:Discussion} contains a discussion of the results and some future directions. Some technical details complementing the calculations from the main text are given in Appendices \ref{app:coeff-table} and \ref{app:double-trace coefficients matching}.

% Preamble:
% \usepackage{tikz}
% \usetikzlibrary{arrows.meta,positioning}

\begin{figure}[t]
\centering
\begin{tikzpicture}[
  font=\small,
  >=Latex,
  box/.style={
    draw, rounded corners=2mm,
    align=center,
    inner xsep=10pt, inner ysep=7pt,
    text width=5.2cm,
    minimum height=12mm
  },
  flow/.style={-Latex, very thick},
  lab/.style={font=\scriptsize, align=center}
]

\node[box] (B1) {Near-boundary expansion};
\node[box, below=12mm of B1] (B2) {Stress tensor coefficients\quad $\Lambda_n$};
\node[box, below=12mm of B2] (B3) {Double-trace coefficients\quad $a_m$};
\node[box, below=12mm of B3] (B4) {Comparison};

\draw[flow] (B1) -- (B2)
  node[midway, right=8mm, lab] {Recurrence\\ relation};

\draw[flow] (B2) -- (B3)
  node[midway, right=8mm, lab] {KMS\\ Pad\'e--Borel};

\draw[flow] (B3) -- (B4)
  node[midway, right=8mm, lab] {Numerical\\ PDE};

\end{tikzpicture}
\caption{Computation flow used in this work.}
\label{fig:computation-flow-stack}
\end{figure}

\bigskip
{\bf Note added:}  As we were finalizing the draft, we became aware of the forthcoming paper \cite{Niarchos:2025cdg} which partially overlaps with our results.

\section{Holographic Thermal Double-Trace Coefficients}
\label{S:Holographic double-trace coefficients}

The aim of this section is to compute the double-trace thermal coefficients appearing in the conformal block decomposition of the $\langle\phi\phi\rangle$ two-point function in holographic CFTs with a large central charge $C_T\gg 1$. To this end, we shall proceed in two steps. First, we obtain thermal coefficients of the stress-tensor sector of the correlator by solving the Klein-Gordon equation on the planar AdS black hole in the near-boundary expansion. While the procedure for computing these coefficients is known, \cite{Fitzpatrick:2019zqz}, we shall reformulate it in a new way as a set of recursion relations which the coefficients satisfy. The recursion relations can be efficiently solved, allowing us to obtain more multi-stress-tensor thermal coefficients than previously available. In Section \ref{SS:Double-trace coefficients from the KMS condition}, we briefly review how the double-trace thermal coefficients are obtained from the stress-tensor coefficients by solving the KMS sum rules and Borel-resumming the solutions, \cite{Buric:2025anb}. In Section \ref{SS:Padé-Borel}, we perform the resummation of the stress-tensor sector numerically using the Pad\'e-Borel method. The numerical summation converges rapidly to yield the double-trace coefficients.

\subsection{Recursion Relations for Multi-Stress Thermal Coefficients}
\label{SS:stress-tensor-recursion}

We study holographic correlators in the state dual to the Euclidean AdS$_5$ planar black hole, described by the metric
\begin{equation}\label{BH-metric}
    ds^2 = r^2f(r)d\tau^2 + \frac{dr^2}{r^2f(r)} + r^2 d\vec{x}^2\,,
\end{equation}
where the function $f(r)$ is given by $f(r) = 1 - \frac{\mu}{r^4}$, $\mu$ is related to the inverse temperature via $\mu = \left(\frac{\pi}{\beta}\right)^4$ and $\Vec{x} = (x^1,x^2,x^3)$ are the usual Cartesian coordinates on $\mathbb{R}^3$. In what follows we set $\beta=1$. In order to compute the two-point correlation function, one solves the Klein-Gordon equation
\begin{equation}\label{PDE}
    \left[\Box - \Delta_\phi(\Delta_\phi-4)\right]\phi(\tau, \vec{x}, r)=0\,,
\end{equation}
for a scalar field $\phi(\tau, \vec{x}, r)$, subject to appropriate boundary conditions\footnote{By a slight abuse of notation we denote the dual CFT operator by $\phi(\tau, \vec{x})$; the distinction should be clear from the context.
}. To study this equation, we perform a near-boundary expansion \cite{Fitzpatrick:2019zqz, Ceplak:2024bja}, introducing a suitable coordinate transformation that isolates the leading asymptotic behavior near the boundary
\begin{equation}
    \rho^2 = r^2\Vec{x}^2, \qquad w^2 = 1 + r^2(\tau^2 + \Vec{x}^2)\ .
\end{equation}
We will be interested in spherically symmetric solutions, $\phi = \phi(r,\rho,w)$. Further, let us factor out the solution in the pure AdS space by putting
\begin{equation}
    \phi(r,\rho,w) = \left(\frac{r}{w^2}\right)^{\Delta_\phi} \psi(r,\rho,w)\ .
\end{equation}
The KG equation then reduces to 
\begin{equation}\label{D-eq}
    D\psi(\tau,\rho,w) = 0\,,
\end{equation}
where a compact expression for the differential operator $D$ can be found in \cite{Ceplak:2024bja}, equations (B.4)--(B.14).
The boundary finite temperature correlator $ g(\tau,\Vec{x}) = \langle \phi(\tau, \vec{x}) \phi(0) \rangle_\beta$ is determined by taking the appropriate limit,
\begin{equation}
    g(\tau,\Vec{x}) = \frac{1}{(\tau^2 + \Vec{x}^2)^{\Delta_\phi}}\lim_{r\to\infty}\psi\ .
\end{equation}
We focus on the stress-tensor sector of the correlator. It was argued in \cite{Fitzpatrick:2019zqz} that the $T_{\mu\nu}^n$ operators probe the bulk geometry near the boundary and one can use  the following ansatz for the stress-tensor sector:
\begin{equation}\label{psi}
    \psi_T =  \sum_{i=0}^\infty \sum_{j=0}^i\sum_{k=-i}^{2i-j} a^i_{j,k} \frac{\rho^{2j} w^{2k}} {r^{4i}}\,, \qquad r\to\infty\,,
\end{equation}
where $a^0_{0,0} = 1$. This large-$r$ expansion on the bulk side corresponds to the OPE of the boundary correlator, which, for $\Vec{x} = 0$, reads
\begin{equation}\label{stress-tensor_g}
    g_T(\tau) = \frac{1}{|\tau|^{2\Delta_\phi}}\sum_{n=0}^\infty \Lambda_n \tau^{4n}\,,
\end{equation}
with $\Lambda_n \equiv a^n_{0,2n}$.
\smallskip

It was shown in \cite{Fitzpatrick:2019zqz} that the form of the solution \eqref{psi} is preserved by the action of the Laplace operator and that the coefficients $a^i_{j,k}$ may be computed by working order by order in $r^{-1}$. However, for $i\gg1$, obtaining the coefficients $a^i_{j,k}$, and thus the stress-tensor data $\Lambda_n$ becomes increasingly time consuming. In order to improve efficiency, we will recast the differential equation \eqref{PDE} as a set of closed form algebraic relations on the coefficients $a^i_{j,k}$. Our method is inspired by a similar treatment of Casimir differential equations satisfied by conformal blocks, \cite{Hogervorst:2013sma,Costa:2016xah}. To proceed, we define
\begin{equation}
    f_{i,j,k} = \frac{\rho^{2j} w^{2k}} {r^{4i}}\ .
\end{equation}
It is convenient to rescale the operator \eqref{D-eq}  $\widetilde{D}$ to
\begin{equation}
     \widetilde{D}\equiv \frac{\left(\pi^4-r^4\right)^2\rho^2  w^4}{r^6}D\ ,
\end{equation}
so that the original Klein-Gordon equation \eqref{PDE} is equivalent to
\begin{equation}
     \widetilde{D}\psi_T(r,\rho, w) = 0\ ,
\end{equation}
and the action of $\widetilde{D}$ on $f_{i,j,k}$ can be written in a compact form
\begin{equation}\label{Df}
   \widetilde{D} f_{i,j,k}(r,\rho,w) = \sum_{(a, b, c) \in S} C_{a,b,c}^{i, j, k} \cdot f_{i + a, j + b, k + c}\ .
\end{equation}
Here 
\begin{align}
    \nonumber S = \{ &(0,0,2),\ (0,1,1),\ (0,1,2),\ (1,0,2),\ (1,1,0),\ (1,1,1),
    \\ &(1,1,2),\ (1,2,0),\ (2,1,0),\ (2,1,1),\ (2,1,2) \}\,,
\end{align}
and the coefficients $C_{a,b,c}^{i, j, k}$, which depend on $i, j, k, a, b, c$ and $\Delta_\phi$, are given in Appendix \ref{app:coeff-table}. Upon summing over the indices $i,j,k$, we can write
 \begin{equation}
     \sum_{i=0}^\infty \sum_{j=0}^i \sum_{k=-i}^{2i-j}\sum_{(a, b, c) \in S} a^i_{j,k}\cdot C_{a,b,c}^{i, j, k} \cdot f_{i + a, j + b, k + c} = 0\ .
 \end{equation}
Setting $I = i + a,\  J = j + b,\ K = k + c$ and substituting these new indices into the sum results in the equation
\begin{equation}
    \sum_{(a, b, c) \in S}\sum_{I,J,K} a^{I-a}_{J-b,K-c}\cdot C_{a,b,c}^{I-a, J-b, K-c} \cdot f_{I, J, K} = 0\ .
\end{equation}
Since the functions \( f_{I, J, K} \) are linearly independent, we obtain a system of constraints on the coefficients
\begin{equation}\label{rec_eq}
\sum_{(a, b, c) \in S} a^{I - a}_{J - b, K - c} \cdot C_{a,b,c}^{I - a, J - b, K - c} = 0\,,
\qquad \text{for all } (I, J, K)\ .
\end{equation}
To construct the recursion procedure, we select the point $(0,1,1)\in S$ and use the last equation to express $a^I_{J-1,K-1}$ in terms of the other ten coefficients entering \eqref{rec_eq}:
\begin{equation}\label{rec_rel}
    a^{I}_{J-1,K-1} = -\frac{1}{C_{0,1,1}^{I - 0, J - 1, K - 1}}\sum_{(a, b, c) \in S\setminus \{(0,1,1)\}} 
a^{I - a}_{J - b, K - c} \cdot 
C_{a,b,c}^{I - a, J - b, K - c}\ .
\end{equation}
We employ this recursion relation to determine all coefficients $a^I_{J,K}$. For this to work, one needs to compute the coefficients in the appropriate order and with appropriate boundary conditions, which ensure that at each step of the procedure, the coefficients on the right hand side of \eqref{rec_rel} have already been computed. The boundary conditions read
\begin{align}
    a^0_{0,0} &= 1 \\
    \nonumber a^I_{J,K} &= 0 \quad \forall\ (I,J,K) \quad \text{not satisfying} \quad I\geq 1,\ 0 \leq J \leq I,\ -I \leq K \leq 2I - J \ .
\end{align}
Although these boundary conditions are formally overdetermined, this will not affect our computation. The algorithm proceeds as follows:
\begin{itemize}
    \item For each \( I \geq 1 \),
    \item For each \( K \in [-I,\, 2I] \),
    \item For each \( J \in [0,\, \min(\, 2I - K,\, I)] \),
    \item Compute \( a^I_{J,K} \) using the recursion relation \eqref{rec_rel}\ .
\end{itemize}
We note again that the quantities of interest are given by $\Lambda_n = a^n_{0,2n}$. The computational efficiency of the procedure allows for routine calculation of exact coefficients of order up to $i \approx 200-300$ without significant computational cost (for a fixed value of $\Delta_\phi$). The coefficients for a general $\Delta_\phi$ are available up to $N=40$.

\subsection{Double-Trace Coefficients from the KMS Condition}
\label{SS:Double-trace coefficients from the KMS condition}

In thermal quantum field theory, the effects of finite temperature can be described by compactifying the theory on a circle of circumference \(\beta = 1/T\). In this work, we consider the theory on the manifold \(S^1_\beta\times\mathbb{R}^3\), the circle factor being parameterized by the Euclidean time $\tau\sim\tau+\beta$. Bosonic fields are required to satisfy periodic boundary conditions along the thermal circle. This leads to the KMS condition, \cite{Kubo:1957mj,Martin:1959jp}, satisfied by scalar two-point functions
\begin{equation}
    g_\beta(\tau,\vec{x}) \equiv
    \langle \phi(\tau, \vec{x})\, \phi(0, 0) \rangle_\beta = \langle \phi(\beta-\tau, \vec{x})\, \phi(0,0) \rangle_\beta\ .
\end{equation}
We shall work at zero spatial separation, $\vec{x} = 0$, and denote the corresponding thermal two-point function by $g_\beta(\tau)$. The idea that we shall follow \cite{El-Showk:2011yvt,Iliesiu:2018fao,Marchetto:2023xap,Buric:2025anb} is to combine the conformal block decomposition of the two-point function and the KMS condition to derive constrains on CFT data. We use the spin-averaged block decomposition
\begin{equation}\label{conformal-block-decomposition-zero-x}
  g_\beta(\tau) = \sum_{\Delta_\mathcal{O}} \frac{a_{\Delta_\mathcal{O}}}{\beta^{\Delta_{\mathcal{O}}}} \tau^{\Delta_{\mathcal{O}}-2\Delta_\phi}\,, \qquad a_\Delta \equiv \sum_{\mathcal{O} \in \phi \times \phi}^{\Delta_\mathcal{O}=\Delta} \frac{\ell_\mathcal{O}+1}{2^{\ell_{\mathcal{O}}}}\lambda_{\phi\phi\mathcal{O}}b_\mathcal{O}\,,
\end{equation}
which can be read off from the correlator at $\vec{x}=0$. Then, the KMS condition leads to the following sum rules
\begin{equation}\label{hol_sum_rules}
    \sum_{m > k} \frac{{a_{[\phi\phi]}}_{2\Delta_\phi+2m}}{2^{2\Delta_\phi+2m}}   \frac{\Gamma(2m+1)}{\Gamma(2(m-k))}  =  - \sum_{n=0
}^{\infty} \frac{\Lambda_n}{2^{4n}}   \frac{\Gamma(4n - 2\Delta_\phi+1)}{\Gamma(4n-2\Delta_\phi-2k)}\,,
\end{equation}
for the thermal coefficients $a_{\Delta_\mathcal{O}}$. We have written the sum rules \eqref{hol_sum_rules} in the form that they take for holographic CFTs. That is, the \(\phi \times \phi\) OPE consists solely of the identity, double-trace, and multi-stress-tensor operators:
\begin{equation}\label{holographic-spectrum}
    \phi \times \phi \sim \mathbbm{1} + [\phi \phi]_{2\Delta_\phi + 2m} + [T_{\mu \nu}]^n\,,
\end{equation}
which holds at leading order in $1/C_T$. The identity contribution is taken into account by including the $n=0$ term on the right-hand side, with $\Lambda_0 = 1$. In Section \ref{SS:stress-tensor-recursion}, we showed how to efficiently compute a large number of stress-tensor coefficients, although a closed-form formula for them remains unknown. In \cite{Buric:2025anb}, it was shown that in the {\it asymptotic approximation}, where $\Lambda_n$ are replaced by their leading large-$n$ asymptotics,  the KMS sum rules can be solved; the solution involves Borel resummation. In the present work, we aim to go beyond the asymptotic approximation by using the exact stress-tensor data rather than their large-$n$ asymptotics.  In \cite{Buric:2025anb} we obtained a solution to the finite version of sum rule equations, that is, to \eqref{hol_sum_rules} where the series on the right-hand side is truncated at order $N$. We denote by $F_k$ the right-hand side of the $k$-th equation with the sum on the right truncated at order $N$,
\begin{equation}\label{finite-sum-rule-corrections}
    F_k = \sum_{n=0}^N  F_k^{(n)} \qquad\text{where}\qquad  F_k^{(n)} = - \frac{ \Lambda_n}{2^{4n}}\frac{\Gamma(4n - 2\Delta_\phi+1)}{\Gamma(4n - 2\Delta_\phi - 2k)} \ .
\end{equation}
Then, according to \cite{Buric:2025anb} (see equation (3.66) of that paper), the double-trace coefficients can be expressed in terms of the stress-tensor coefficients as
\begin{align}
    \nonumber a_m = &\sum_{n=0}^{\lfloor\Delta_\phi/2 \rfloor}2^{-2m-2\Delta_\phi+1} \Lambda_n\,\zeta (2m + 2\Delta_\phi - 4n) (2\Delta_\phi-4n)_{2m}\ \\
     + &\sum_{n=\lfloor\Delta_\phi/2\rfloor+1}^N \sum_{k=0}^{2n-m-\lfloor\Delta_\phi\rfloor} (-1)^k \frac{\zeta(2k)}{\pi^{2k}} \left(2-4^{1-k}\right)F_{k+m-1}^{(n)}\,, \label{Borel_sol}
\end{align}
where the sum in the second line is set to zero, whenever $2n-m-\lfloor\Delta_\phi\rfloor<0$. In the last line, we have introduced the coefficients $a_m$ which are related to original double-trace double-trace coefficients as
\begin{equation}\label{notations}
    {a_{[\phi\phi]}}_{2\Delta_\phi+2m}= \frac{2^{2\Delta_\phi+2m}\,a_m}{(2m)!}\,,
\end{equation}
as this simplifies several formulas in what follows. The identity \eqref{Borel_sol} was derived by first solving the KMS equations of the form \eqref{hol_sum_rules} for each stress-tensor block $F_k^{(n)}$ individually, then applying Borel resummation to the resulting expressions, and finally summing over $n$.\footnote{The need for Borel resummation in a broadly similar context was also noticed in \cite{Karydas:2023ufs}.} Note that the contribution from each block coincides with the double-trace coefficient in the GFF with conformal dimension $\Delta_\phi - 2n$, up to an overall factor $\Lambda_n$.

\subsection{Padé-Borel Summation}
\label{SS:Padé-Borel}

Even though the expression for double-trace coefficients in \eqref{Borel_sol} naturally splits into two parts over $n$ -- from 0 to $\lfloor\Delta_\phi/2 \rfloor$ and from $\lfloor\Delta_\phi/2 \rfloor + 1$ to $N$ -- we will write it uniformly as 
\begin{equation}\label{am-divergent-series}
   a_m \equiv a_m^{[N]} = \sum_{n=0}^N a_m^{(n)}\ .
\end{equation}

As $n\to\infty$, the coefficients $a^{(n)}_m$ grow as $a^{(n)}_m \sim (4n)!$, see \cite{Buric:2025anb}. Therefore, for any fixed $m$, the sum \eqref{am-divergent-series} diverges in the limit $N \to \infty$. To address this issue, we employ the Pad\'e approximation of the Borel-transformed series.\footnote{This Borel resummation refers to the series over  $n$, and should not be confused with the earlier Borel resummation applied to each $F_k^{(n)}$ individually.} In addition, to improve the numerical accuracy, we perform the inverse Borel transform with the help of Gauss-Laguerre quadrature. The full numerical method (proposed in \cite{RAZAFINDRALANDY201356} for more general classes of problems) is summarized in Table \ref{BP_table}. Let us give more details\footnote{We assume that $N$ is even. A similar construction can be repeated for odd $N$.}.\par
\begin{table}[ht]\centering  {\sf\small
\begin{tabular}{ccccc}
	$a_m^{[N]}(z) = \sum_{n=0}^N a_m^{(n)} z^n$&  &\hspace{-0.7cm}$a_m^{[N,N_G],\text{PB}} =\sum_{i = 1}^{N_G}\mathcal{P}^{[N]}(t_i)\omega_i$ 
\\
 $\left.\begin{array}{c}\\\text{Borel}\\\\\end{array}\right\downarrow$&
 &$\left\uparrow \begin{array}{c}\\\text{Gauss-Laguerre} \\\\\end{array}\right.$
\\
$
 \mathcal{B} a_m^{[N]}(t^4) = \sum_{n=0}^N \frac{ a_m^{(n)}}{(4n)!}t^{4n}$
&$\overrightarrow{\hspace{.5cm}\text{ Padé}\hspace{.5cm}}$
&$\mathcal{P}^{[N]}(t) = \frac{p_0 + p_1t^4 + \ldots + p_{N/2}t^{2N}}{q_0 + q_1t^4 + \ldots + q_{N/2}t^{2N}}$\\\\
\end{tabular}}
\caption{Padé-Borel summation algorithm}\label{BP_table}
\end{table}

\begin{figure}
    \centering
    \includegraphics[width=1\linewidth]{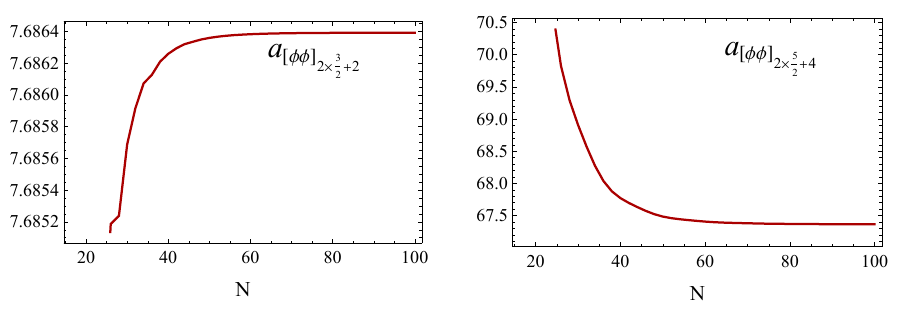}
    \caption{Padé-Borel approximations of the first ($m=1$) and second ($m=2$)  double-trace coefficients $a_{[\phi\phi]_{2\Delta_\phi + 2m}}$ for $\Delta_\phi = 3/2$ and $\Delta_\phi = 5/2$, respectively, as functions of the number $N$ of the included stress-tensor coefficients.}
    \label{fig:Pade-Borel_dt}
\end{figure}

To begin with, we promote $a_m^{[N]}$ to a function of a variable $z$ by
\begin{equation}
   a_m^{[N]}(z) = \sum_{n=0}^N a_m^{(n)} z^n\,,
\end{equation}
and perform the generalized Borel transform of the latter,
\begin{equation}\label{truncated-Borel}
    \mathcal{B} a_m^{[N]}(t^4) = \sum_{n=0}^N \frac{ a_m^{(n)}}{(4n)!}t^{4n}\ .
\end{equation}
Next, we construct a Padé approximant to \eqref{truncated-Borel}. We will use a diagonal Padé approximant, which replaces the polynomial \eqref{truncated-Borel} by a rational function of the form
\begin{equation}
    \mathcal{P}^{[N]}(t) = \frac{P_{N/2}(t^4)}{Q_{N/2}(t^4)}\ .
\end{equation}
Here, $P_{N/2}$ and $Q_{N/2}$ are polynomials of degree $N/2$, chosen such that the Taylor expansion of $\mathcal{P}^{[N]}(t)$ up to the order $t^{4N}$ is given by \eqref{truncated-Borel}. The Pad\'e-Borel approximated value of the original sum \eqref{am-divergent-series} is then obtained via the Laplace transform
\begin{equation}\label{Laplace-transform}
    a_m^{[N],\text{PB}} = \int_0^\infty e^{-t} \mathcal{P}^{[N]}(t)\, dt\ .
\end{equation}
In practice, we perform the integral numerically using the Gauss-Laguerre quadrature. This gives
\begin{equation}\label{PB-sum}
    a_m^{[N,N_G],\text{PB}} = \sum_{i = 1}^{N_G}\mathcal{P}^{[N]}(t_i)\,\omega_i\,,
\end{equation}
where $t_i$ is the $i$-th root of the Laguerre polynomial $L_{N_G}(t)$ of order $N_G$, and associated weights $\omega_i$ are given by (see \cite{Abramowitz:1964:HMF})
\begin{equation}
\omega_i = \frac{t_i}{(N_G+1)^2[L_{N_G+1}(t_i)]^2}\ .    
\end{equation}
As we perform the resummation \eqref{PB-sum} for increasing values of $N$ (and $N_G$), the result converges to a finite value. The values of successive Pad\'e-Borel approximations to the double-trace coefficients are shown in Figure \ref{fig:Pade-Borel_dt}. In Table \ref{table-PB-coefficients-32}, we provide the thermal coefficients of low-lying double-trace operators, with $\Delta_\phi=3/2$, keeping the number of digits that are stable under the increase of $N$ and $N_G$. These double-trace coefficients are the main result of the present section. The same procedure can be repeated for any value of $\Delta_\phi$. We display the resulting low-lying double-trace coefficients for $\Delta_\phi=5/2$ in the next section - see Table \ref{table-results-52-2}. Let us remark that the first double-trace coefficient, corresponding to the operator $\phi^2$, is not constrained by the KMS condition and hence does not appear in the table. Indeed, in the conformal block decomposition \eqref{conformal-block-decomposition-zero-x}, this coefficient is multiplied by a constant function, which is KMS-invariant on its own.

\begin{table}
\centering
\begin{tabular}{ | c | c |}
 \hline
  operator $\mathcal{O}$ & Pad\'e-Borel resummed coefficient $a_\mathcal{O}$ \\ 
 \hline
 $[\phi\phi]_{2\Delta_\phi+2}$ & 7.686391301178 \\
 \hline
 $[\phi\phi]_{2\Delta_\phi+4}$ & 38.28355969692 \\
 \hline
 $[\phi\phi]_{2\Delta_\phi+6}$ & 72.763467218 \\
 \hline
 $[\phi\phi]_{2\Delta_\phi+8}$ & 56.85578514 \\
 \hline
 $[\phi\phi]_{2\Delta_\phi+10}$ & 66.293380 \\
 \hline
 $[\phi\phi]_{2\Delta_\phi+12}$ & 312.632 \\
 \hline
 $[\phi\phi]_{2\Delta_\phi+14}$ & 499.994 \\
 \hline
 $[\phi\phi]_{2\Delta_\phi+16}$ & -212.0 \\
 \hline
\end{tabular}

\caption{Double-trace coefficients obtained via Pad\'e-Borel resummation for $\Delta_\phi = \frac32$.}

\label{table-PB-coefficients-32}
\end{table}

\subsection{Bound on the Holographic Correlator from its Real-Line Value}
\label{SS:ConsistencyChecks}

In this section, we examine the consistency of the two-point function \({g}_\beta(\tau)\), constructed from the stress-tensor coefficients provided in \eqref{stress-tensor_g} and the double-trace coefficients in \eqref{Laplace-transform}, with general expectations from conformal field theory at finite temperature, as well as with known features of thermal correlators in holographic CFTs. By construction, \({g}_\beta(\tau)\) is a KMS-invariant function that admits an expansion in conformal blocks \eqref{conformal-block-decomposition-zero-x} consistent with the holographic spectrum \eqref{holographic-spectrum}. We now investigate whether the thermal two-point function $g_\beta(\tau)$ satisfies the boundedness condition in the complex $\tau$-plane (see e.g. \cite{Buric:2025anb, Fidkowski:2003nf}).

\begin{figure}[ht]\label{fig:boundedness}

\begin{subfigure}[t]{.49\textwidth}
  \centering
  \includegraphics[width=\textwidth]{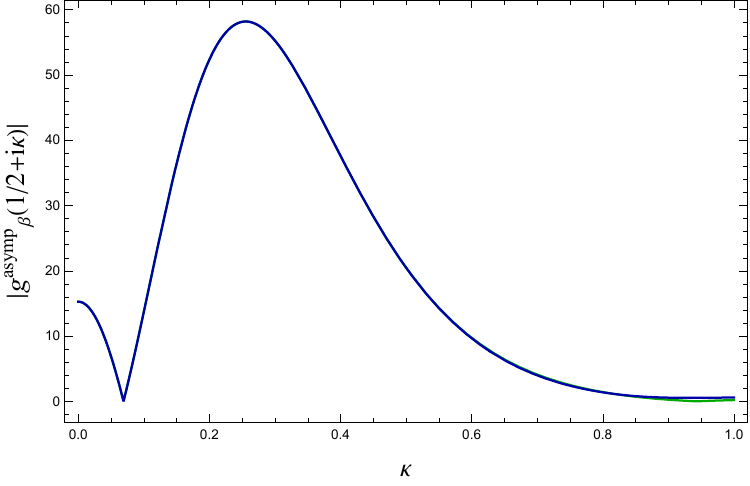}
  \caption{}
  \label{fig:boundedness_asymp}
\end{subfigure}%
\hfill
\begin{subfigure}[t]{.49\textwidth}
  \centering
  \includegraphics[width=\linewidth]{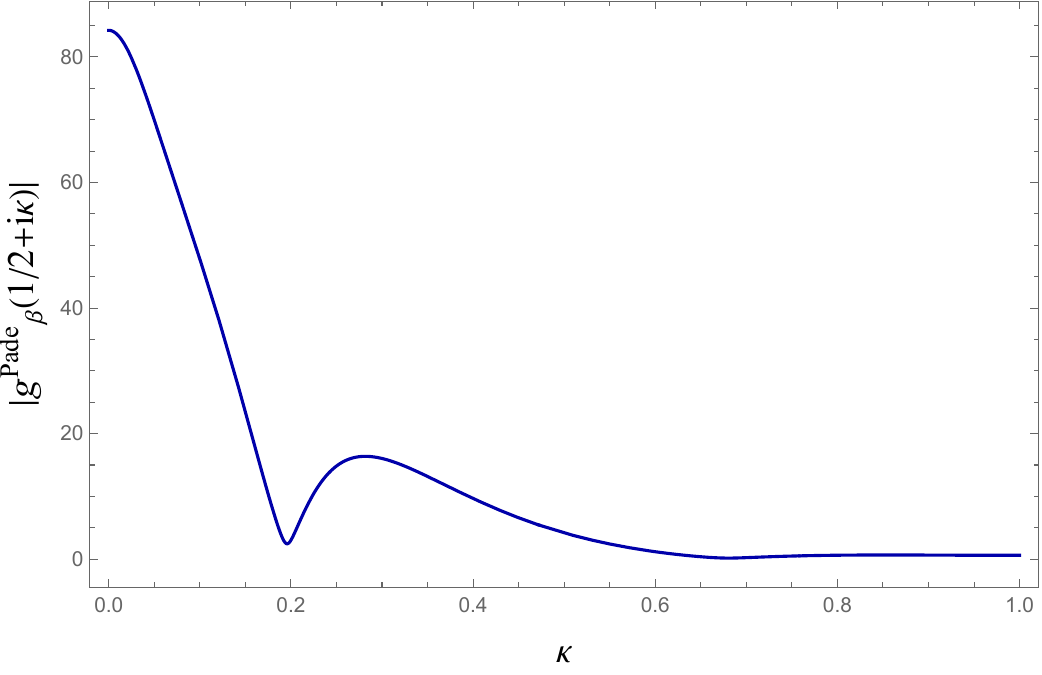}
  \caption{}
  \label{fig:boundedness_real}
\end{subfigure}
\caption{Verifying the boundedness condition \eqref{bound}. The blue curve in \textbf{plot (a)} on the left shows the diagonal Padé approximation  for the two-point function in the \emph{asymptotic model}, constructed from the first 15 conformal blocks with conformal dimension $\Delta_\phi = \frac52$, plotted as a function of $\kappa$ ($\tau = \frac12 + i\kappa$). The green curve represents the exact two-point function.
The \textbf{right plot (b)} shows the diagonal Padé approximation $g^{\text{Padé}}_\beta(\tau)$ to the full holographic correlator constructed from the first 15 conformal blocks of \( g_\beta(\tau) \) with conformal dimension \(\Delta_\phi = \frac52\). For both plots, $\beta$ is taken to be equal to 1.}
\end{figure}

\paragraph{Boundedness in the Complex Plane.}
We test the boundedness condition,
\begin{equation}\label{bound}
    g_\beta(\tau) \geq |g_\beta(\tau + i \kappa)|\,, \qquad \tau,\kappa\in\mathbb{R}\ .
\end{equation}
This property is particularly noteworthy, as our previous paper \cite{Buric:2025anb} showed that the asymptotic model\footnote{Here, the \emph{asymptotic model} refers to the construction in which the two-point function is obtained by solving the KMS condition using the large-$n$ asymptotic expressions for the stress-tensor coefficients $\Lambda_n$.} may fail to satisfy the boundedness condition, depending on the value of $\Delta_\phi$ (see Figure \ref{fig:boundedness_asymp}). This failure arises from the condition’s sensitivity to the low-lying stress-tensor and double-trace coefficients and highlights the inability of the asymptotic model to accurately capture  relevant physics. 
While the two-point function in the asymptotic model can be obtained explicitly (see Appendix A.1 of \cite{Buric:2025anb}), 
the Padé-Borel method provides only low-lying thermal coefficients, as higher-order coefficients require a larger number $N$ of stress-tensor inputs to converge reliably (see Figure \ref{fig:Pade-Borel_dt}). 
We approximate the holographic the two-point function by constructing a Pad\'e approximant $g^{\text{Padé}}_\beta(\tau)$ of the spin-averaged block decomposition \eqref{conformal-block-decomposition-zero-x} with only several known blocks.
The result is robust under variations in the number of conformal blocks and the Pad\'e approximation order. We have also applied the same method to the asymptotic model to verify that the Pad\'e-approximated result closely matches the exact two-point function (compare the blue and green lines in Figure \ref{fig:boundedness_asymp}, which are on top of one another.
The result is shown in Figure \ref{fig:boundedness_real} - we see that \eqref{bound} is indeed satisfied in the full holographic correlator, which provides another consistency check of our method.

\section{Comparison with PDE Solution}
\label{S:Matching to the PDE solution}

In this section, we approach the computation of the holographic thermal double-trace coefficients from the bulk perspective. In Section \ref{SS:PDE}, we review how these coefficients may be computed by numerically solving the partial differential equation (PDE) which follows from the Klein-Gordon equation on the planar AdS black hole background. In Section \ref{SS:Matching}, we apply this method and compare the resulting double-trace coefficients with the ones found in Section \ref{S:Holographic double-trace coefficients}. The results obtained from the two methods show excellent numerical agreement.

\subsection{Numerical PDE Solution}\label{SS:PDE}

The holographic thermal two-point function is extracted from the boundary limit of a solution to the Klein-Gordon equation on the planar AdS-Schwarzschild background. In order to fix the appropriate solution, one needs to specify the boundary conditions both at the asymptotic boundary and the black hole horizon. In Section \ref{SS:stress-tensor-recursion}, we reviewed how the stress-tensor sector of the correlator can be computed from the KG equation. This part of the correlator is only sensitive to the boundary condition at $r\to\infty$ and is fully captured by the ansatz \eqref{psi}.
\smallskip

At present, no analogous ansatz that would capture both the stress-tensor and the double-trace contributions to the correlator is known. Nevertheless, one may proceed by solving the KG equation, with both sets of boundary conditions, numerically. This approach was proposed in \cite{Parisini:2023nbd}, where the authors used it to obtain the value of the first double-trace coefficient $a_{\phi^2}$. We will follow the method of \cite{Parisini:2023nbd} - since the latter was fully described in that work and also reviewed in \cite{Buric:2025anb}, we shall be brief.
\smallskip

The numerical analysis is conveniently done in coordinates $(\rho,\varphi,R)$ given by
\begin{equation}\label{cube-coordinates}
    r^{-1} = 1 - \rho^2\,, \qquad \tau = \frac{\varphi}{2}\,, \qquad |\vec{x}| = \frac{R}{1-R^2}\,,
\end{equation}
in terms of the coordinates used in \eqref{BH-metric}. To match the notation of \cite{Parisini:2023nbd}, we set $\beta = \pi$. The boundary conditions supplementing the KG equation \eqref{PDE} are regularity in the bulk and a delta-function-like behavior at the boundary $\partial=\{r=\infty\}$,
\begin{equation}
    \phi|_\partial = \tilde G_{\text{AdS}}|_\partial\,, \qquad \tilde G_{\text{AdS}} \equiv \frac{\left(1-\rho^2\right)^{\Delta_\phi}}{\left(\left(1-\rho^2\right)^2 + \frac14 F(\varphi) + \frac{R^2}{(1-R^2)^2}\right)^{\Delta_\phi}}\ .
\end{equation}
Here, $F(\varphi)$ is a $2\pi$-periodic function whose Taylor series around $\varphi=0$ starts as $\varphi^2$. If the term $F(\varphi)$ was replaced by $\varphi^2$, the function $\tilde G_{\text{AdS}}$ would reduce to the propagator on the empty AdS space. In practice, the function $F$ is chosen such that the next term in the Taylor series has a high power. As long as this is satisfied, the precise choice of $F(\varphi)$ does not affect the resulting low-lying double-trace coefficients which are extracted from the solution at low $\varphi$.
\smallskip

The coordinate $\varphi$ is periodic with period $2\pi$, while $\rho$ and $R$ take values in the interval $[0,1]$. Therefore, in order to have a well-posed problem, one needs to specify Dirichlet and Neumann boundary conditions on the four loci: $\{\rho=0\}$, $\{\rho=1\}$, $\{R=0\}$ and $\{R=1\}$. The appropriate conditions are conveniently stated for the function $H$ related to the KG mode $\phi$ via
\begin{equation}
    H = (1-\rho^2)^{-\Delta_\phi} \left(\phi - \tilde G_{\text{AdS}}\right)\ .
\end{equation}
The correct choice of boundary conditions depends on the value of $\Delta_\phi$. From this point on, we focus on the case $\Delta_\phi=3/2$. Then the conditions read
\begin{equation}\label{Dirichlet-and-Neumann}
    H|_{R=1} = \partial_R H |_{R=0} =\partial_\rho H |_{\rho=0} = \partial_\rho H |_{\rho=1} = 0\ .
\end{equation}
Together with periodicity in $\varphi$, these boundary conditions give a well-posed problem. The latter may be solved numerically. Once the solution for $H$ has been obtained, the thermal two-point function is extracted from it via
\begin{equation}\label{H-gbeta-relation}
    g_\beta(\tau,R) = H(\rho=1,\varphi,R) + \frac{1}{\left(\frac14 F(\varphi) + \frac{R^2}{(1-R^2)^2}\right)^{\Delta_\phi}}\ .
\end{equation}
The numerical procedure that we follow is discretization of the PDE satisfied by the function $H$ on a grid consisting of $L^3$ points, with $L=170$. The distribution of nodes of the grid is uniform in directions of $\varphi$ and $R$ and follows the Chebyshev collocation in the $\rho$-direction.

\subsection{Pad\'e-Borel vs.  PDE Results}
\label{SS:Matching}

After extracting the two-point function $g_\beta(\tau)$\footnote{Note that to align with the notation used in \cite{Parisini:2023nbd}, in this section we set $\beta = \pi$.} from the asymptotic expansion of the numerical solution, one can compare it against the thermal OPE. This OPE includes the stress-tensor coefficients computed in Section \ref{SS:stress-tensor-recursion} and the double-trace coefficients derived in Section \ref{SS:Padé-Borel}. However, the comparison is not entirely straightforward, as the region near $\tau = 0$ cannot be directly used: the two-point function develops a $1/|\tau|^{2\Delta_\phi}$ singularity at this point, making the numerical solution unstable.
\smallskip

\begin{figure}[ht]\label{fig:PDE_der_comparison}

\begin{subfigure}[t]{.49\textwidth}
  \centering
  \includegraphics[width=\textwidth]{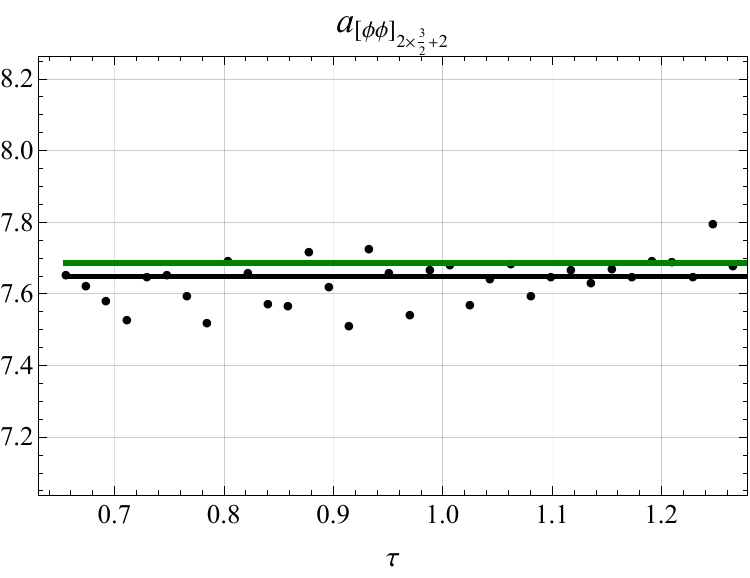}
  \caption{}
  \label{fig:PDE_der_comparison_32s}
\end{subfigure}%
\hfill
\begin{subfigure}[t]{.49\textwidth}
  \centering
  \includegraphics[width=\linewidth]{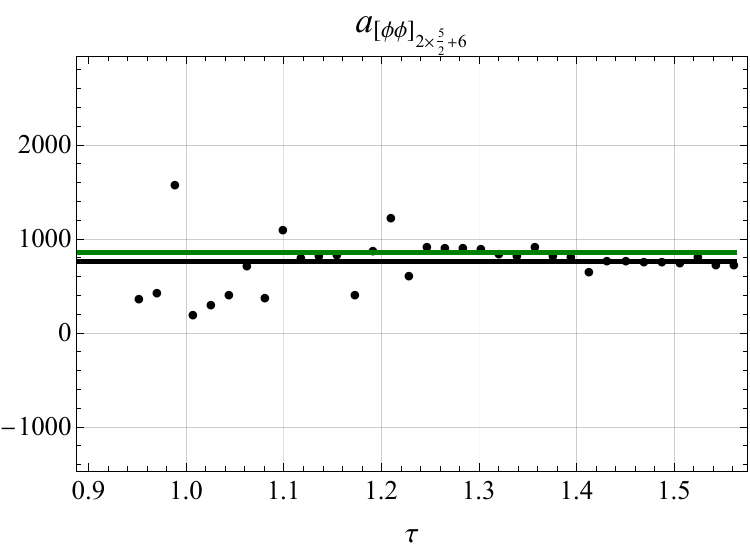}
  \caption{}
  \label{fig:PDE_der_comparison_52s}
\end{subfigure}
\caption{Comparing PDE and Pad\'e-Borel results for the double-trace coefficients. The \textbf{left picture (a)} pertains to the double-trace coefficient $a_{[\phi\phi]_{2\Delta_\phi+2}}$ in the OPE of two scalar fields of conformal dimension $\Delta_\phi = \frac32$. The black dots represent the values of $a_{[\phi\phi]_{2\Delta_\phi+2}} + o(\tau^{12})$ from \eqref{fit2}, obtained from the PDE solution. The black line highlights the flattest segment of these data points. The green line depicts the corresponding Pad\'e-Borel resummed result. The \textbf{right picture (b)} corresponds to the double-trace coefficient $a_{[\phi\phi]_{2\Delta_\phi+6}}$ with $\Delta_\phi = \frac52$.}
\end{figure}

There are several ways to extract double-trace coefficients from the numerical solution. We shall present one method here and another one in Appendix \ref{app:double-trace coefficients matching} - they lead to similar results. We begin by truncating the thermal OPE and fixing several of the stress-tensor and double-trace coefficients to the values obtained in Section \ref{S:Holographic double-trace coefficients}, leaving the leading double-trace coefficient $a_{[\phi\phi]_{2\Delta_\phi}}$ unfixed. Note that this coefficient is not determined by KMS sum rules and therefore cannot be computed using the methods described in Section \ref{SS:Padé-Borel}. Therefore, we aim to estimate it by comparing the numerical PDE solution with the two-point function with subtracted contributions of several low-lying operators. For instance, in the case $\Delta_\phi = \frac32$
\begin{align}
    \nonumber &g_\beta(\tau) - \frac{1}{\tau^3} - \frac{3\tau}{80}\frac{\pi^4}{\beta^4} - \frac{7.686}{\beta^5}\tau^2 - \frac{38.28}{\beta^7}\tau^4 + \frac{479 \tau^5}{268800}\frac{\pi^8}{\beta^8} - \frac{72.76}{\beta^9} \tau^6 - \frac{56.86}{\beta^{11}} \tau^8 - \\ 
    & \frac{79063 \tau^9}{1107025920}\frac{\pi^{12}}{\beta^{12}} - \frac{66.29}{\beta^{13}}\tau^{10} - \frac{312.6}{\beta^{15}}\tau^{12} + \frac{103234783\tau^{13}}{35670078259200}\frac{\pi^{16}}{\beta^{16}} \approx \frac{a_{[\phi\phi]_{2\Delta_\phi}}}{\beta^{3}} + o(\tau^{14})\ .\label{fit}
\end{align}
The coefficients multiplying powers of $\tau$ in this equation were obtained in Sections \ref{SS:stress-tensor-recursion} and \ref{SS:Padé-Borel} (we display only a first few digits of double-trace coefficients in \eqref{fit} for notational simplicity). We compute the left hand side of \eqref{fit} by substituting for $g_\beta(\tau)$ the numerical PDE solution. For small values of $\tau$, it is observed that this gives approximately a constant function, in agreement with the expectation from the OPE. To estimate the value of the constant, we use the flattest part of the plot of the left hand side.\footnote{While the error term on the right hand side of \eqref{fit} becomes smaller for $\tau\to0$, the instability of the numerical solution forbids us from using very small $\tau$ to estimate $a_{[\phi\phi]_{2\Delta_\phi}}$.} This reproduces the result of \cite{Parisini:2023nbd},
\begin{equation}\label{leading-dt-coefficient}
    a_{[\phi\phi]_{2\Delta_\phi}} \approx 1.10735\ .
\end{equation}
We move to higher double-trace coefficients. To approximate the second coefficient $a_{[\phi\phi]_{2\Delta_\phi+2}}$, we consider the derivative of the correlator with the subtracted leading order terms except $a_{[\phi\phi]_{2\Delta_\phi+2}}$, namely
\begin{align}
    \nonumber & \frac{1}{2\tau}\partial_\tau\bigg(g_\beta(\tau) - \frac{1}{\tau^3} - \frac{a_{[\phi\phi]_{2\Delta_\phi}}}{\beta^3} -\frac{3\tau}{80}\frac{\pi^4}{\beta^4} - \frac{38.28}{\beta^7}\tau^4 + \frac{479 \tau^5}{268800}\frac{\pi^8}{\beta^8} - \frac{72.76}{\beta^9} \tau^6 - \\ \nonumber
    &\frac{56.86}{\beta^{11}} \tau^8 - \frac{79063 \tau^9}{1107025920}\frac{\pi^{12}}{\beta^{12}} - \frac{66.29}{\beta^{13}}\tau^{10} - \frac{312.6}{\beta^{15}}\tau^{12} + \frac{103234783\tau^{13}}{35670078259200}\frac{\pi^{16}}{\beta^{16}}\bigg) \approx \\ & \frac{a_{[\phi\phi]_{2\Delta_\phi+2}}}{\beta^5} + o(\tau^{12})\ .\label{fit2}
\end{align}
Notice that we have taken the derivative with respect to $\tau$ in order to eliminate the dependence of the subtracted two-point function on $a_{[\phi\phi]_{2\Delta_\phi}}$. Again, upon substituting $g_\beta(\tau)$ on the left hand side by the numerical solution, the result is seen to be approximately a constant function - see Figure \ref{fig:PDE_der_comparison_32s}. We read off the estimate for $a_{[\phi\phi]_{2\Delta_\phi+2}}$ from the flattest part of the plot. Higher double-trace coefficients can be obtained similarly (an example is given in Figure \ref{fig:PDE_der_comparison_52s}).
\smallskip

Our results are shown in Table \ref{table-results-32-2}. We see that the relative discrepancy between estimates obtained via the Pad\'e-Borel method and the PDE solution is less than $1\%$ for the first three lowest coefficients $a_{[\phi\phi]_{2\Delta_\phi+2}}$, $a_{[\phi\phi]_{2\Delta_\phi+4}}$, $a_{[\phi\phi]_{2\Delta_\phi+6}}$ and around $6\%$ for the fourth coefficient. The corresponding double-trace coefficients in GFF are shown in the third column - they are clearly separated from the two numerical estimates of holographic coefficients. For completeness, we also include the double-trace coefficients of the asymptotic model of \cite{Buric:2025anb}.

\begin{table}
\centering
\begin{tabular}{ | c | c | c | c | c |}
 \hline
  $\Delta_\phi = \frac32$ & Pad\'e-Borel & PDE & GFF & AM  \\ 
 \hline
 $[\phi\phi]_{2\Delta_\phi+2}$ & 7.686 & 7.7 & 12.44 & 7.549 \\
 \hline
 $[\phi\phi]_{2\Delta_\phi+4}$ & 38.28 & 38 & 30.26 & 37.60 \\
 \hline
 $[\phi\phi]_{2\Delta_\phi+6}$ & 72.76 & 72 & 56.11 & 71.54 \\
 \hline
 $[\phi\phi]_{2\Delta_\phi+8}$ & 56.86 & 50 & 90.04 & 58.91 \\
 \hline
\end{tabular}
\caption{Double-trace coefficients using Pad\'e-Borel and numerical solution to PDE.}
\label{table-results-32-2}
\end{table}

\begin{table}
\centering
\begin{tabular}{ | c | c | c | c | c |}
 \hline
  $\Delta_\phi = \frac52$ & Pad\'e-Borel & PDE & GFF & AM \\ 
 \hline
 $[\phi\phi]_{2\Delta_\phi+2}$ & 37.67 & 40 & 30.25 & -129.7\\
 \hline
 $[\phi\phi]_{2\Delta_\phi+4}$ & 67.36 & 80 & 140.3 & -324.0 \\
 \hline
 $[\phi\phi]_{2\Delta_\phi+6}$ & 851.8 & 800 & 420.2 & 1644 \\
 \hline
 $[\phi\phi]_{2\Delta_\phi+8}$ & 2505 & 2000 & 990.1 & 4487 \\
 \hline
\end{tabular}
\caption{Double-trace coefficients for $\Delta_\phi=5/2$.}
\label{table-results-52-2}
\end{table}

\medskip

We performed a similar analysis for the $\Delta_\phi=5/2$ case. On the numerical PDE side, this requires modifying the boundary conditions appropriately, see \cite{Parisini:2023nbd} for details. The results are summarized in Table \ref{table-results-52-2}.\footnote{For $\Delta_\phi=5/2$, the asymptotic model yields a result which is quite far from the full holographic correlator, as evident from Figure \ref{fig:boundedness_asymp} and Table \ref{table-results-52-2}.}

\section{Discussion}
\label{S:Discussion}

In this work, we have used the KMS condition to determine thermal coefficients of low-lying double-trace operators in holographic CFTs. The method is largely based on our previous work \cite{Buric:2025anb}, where it was noticed that holographic KMS sum rules may be solved explicitly given the stress-tensor part of the two-point function. Importantly, the solution requires Borel resummation in order to be made convergent. Since only a finite number of thermal coefficients from the stress-tensor sector is available, exact Borel resummation is not possible. In this paper, we followed a variation of this method by truncating the block expansion of the stress-tensor sector at order $N$ and using both Borel resummation and a Pad\'e approximation. As $N$ is increased, this Pad\'e-Borel method produces rapidly convergent thermal coefficients of the double-trace sector. E.g. for the external scalar of dimension $\Delta_\phi = 3/2$, the resulting low-lying double-trace coefficients, with the number of stable digits are shown in Table \ref{table-PB-coefficients-32}.
\smallskip

The same double-trace data is expected to be contained in solutions to the Klein-Gordon equation on the planar AdS-Schwarzschild black hole. In order to verify this, we followed the numerical procedure for solving this equation proposed in \cite{Parisini:2023nbd}, supplementing it with some technical additions that were needed to increase the accuracy. With this improvement, we were able to compare the results with the Pad\'e-Borel resummation - the methods show excellent agreement. E.g. for $\Delta_\phi=3/2$, the two estimates of the first three double-trace coefficients agree to within $1\%$. The results of the comparison are summarized in Tables \ref{table-results-32-2} and \ref{table-results-52-2}.
\smallskip

Let us note that accurate determination of double-trace coefficients using the bulk PDE is a difficult numerical problem. This is especially true for higher coefficients, whose values are increasingly sensitive to the numerical solution. By comparison, the Pad\'e-Borel method is much more stable (as well as efficient). The number of double-trace coefficients that can be accurately predicted by the latter depends on the number $N$ of available stress-tensor coefficients, which are used as the input. In turn, the stress-tensor coefficients are found by solving the KG equation on the black hole background in the near-boundary expansion, \cite{Fitzpatrick:2019zqz}.
%\footnote{Note that this solution differs from the numerical solution because of boundary conditions.} 
We also developed a set of recursion relations for the coefficients in this expansion. The recursion relations significantly improve the efficiency with which these coefficients are computed and may be of interest in their own right. In this work, we used $N=300$ coefficients for $\Delta_\phi = 3/2$ and $N=200$ for $\Delta_\phi=5/2$.
\medskip

We take the above results as a demonstration that the holographic KMS sum rules are efficiently solved by the Pad\'e-Borel method to give low-lying double-trace CFT data. This opens several interesting avenues for further studies.

Firstly, we note that the low-lying double-trace coefficients found in this work complement the asymptotic analysis of \cite{Buric:2025anb}. Taken together, these two results give precise estimates of the thermal coefficients for all the operators appearing in the $\phi\times\phi$ OPE. It would be interesting to extract from the resulting two-point function Lanczos coefficients and quasi-normal modes (see \cite{Dymarsky:2021bjq,Nandy:2024evd,Dodelson:2024atp,Dodelson:2025rng,Rabinovici:2025otw,Demulder:2025uda} for some recent papers on this topic).
\smallskip

An important extension of the method that we will study in the future is to non-zero spatial separation $\vec{x}$. This would allow one to probe thermal coefficients of double-trace operators distinguished by the spin, $a_{[\phi\phi]_{n,\ell}}$, as opposed to the spin-averaged coefficients $a_{[\phi\phi]_{2\Delta_\phi+2m}}$ studied above. The thermal coefficients of the stress-tensor sector necessary for this analysis are already known - thus, what remains is to generalize the solutions to KMS sum rules to $\vec{x}\neq0$. A more challenging extension is to CFTs on $S^1 \times S^{d-1}$. From the gravity side, the near-boundary expansion is also available in the case of the spherical AdS black hole, \cite{Fitzpatrick:2019zqz}.\footnote{For exact approaches to the KG equation on the AdS-Kerr black hole through relations to supersymmetric gauge theories, see \cite{Dodelson:2022yvn,Aminov:2023jve} and references therein. The main difference from the present work is that here we compute the Euclidean thermal correlator in position space, $g_\beta(t,\mathbf{x}=0)$, whereas \cite{Dodelson:2022yvn,Aminov:2023jve} focus on the retarded correlator in momentum space, $G_R(\omega,\mathbf{k})$; the relation between the two is reviewed in Appendix A of \cite{Dodelson:2022yvn}. It would be interesting to use their exact thermal two-point function to study double-trace contributions in an instanton expansion.} On the other hand, the kinematics of CFT correlators on the geometry $S^1\times S^{d-1}$ is starting to be explored, \cite{Gobeil:2018fzy,Alkalaev:2022kal,Alkalaev:2023evp,Alkalaev:2024jxh,Buric:2024kxo,Buric:2025uqt,Ammon:2025cdz}.
\smallskip

Another natural question that arises is what other CFTs, beyond holography, are amenable to the Pad\'e-Borel technique. The question seems to be related to the following: {\it What CFT data are sufficient to uniquely specify a thermal two-point function?} The present paper and previous work \cite{Buric:2025anb} give strong evidence that for holographic theories at large $C_T$, the stress-tensor sector of the correlator is sufficient. It is known that the KMS condition together with a small amount of low-lying CFT data constrains thermal two-point functions in the Ising and $O(N)$ models, \cite{Iliesiu:2018fao,Iliesiu:2018zlz,Barrat:2025wbi}, see also \cite{Petkou:1998fb}. Can the method of this work be adapted to these models to give high precision estimates of thermal coefficients? The situation in the large-$N$ limit of the $O(N)$ model, is particularly interesting, as the spectrum of the theory is very similar to that of a holographic CFT. Namely, in $d=3$ there are two towers of operators -- $\sigma^n$ with conformal dimensions $2 n$ and $[\phi\phi]_{n,\ell}$ with conformal dimensions $2 n +\ell +1$. The former are the analogs of the multi-stress operators, while the latter are the analogs of the double traces.\footnote{An important difference between the $O(N)$ model and holography is the behavior of  $\sigma^n$ (in the $O(N)$ model) and multi-stress (in holography) thermal coefficients. The former decay as the conformal dimension grows, but  the latter exhibit exponential growth related to the bouncing singularity.} Hence, we expect that both in the large $N$ limit of the $O(N)$ model and in holography one needs to input the thermal coefficients of one tower to fix the other with the KMS condition.

\section*{Acknowledgements}

We wish to thank J. Barrat, D. Bozkurt, N. \v Ceplak, C. Esper, E. Helfenberger, Z. Komarkodski, M. Kulaxizi, H. Liu, E. Marchetto, A. Miscioscia, V. Niarchos, C. Papageorgakis, E. Pomoni, F. Russo, K. Skenderis, A. Stergiou, S. Valach, B. Withers  for discussions, correspondence and comments on the manuscript. 
A.P. thanks the Simons Center for Geometry and Physics (Stony Brook) for its hospitality during the program 'Black Hole Physics from Strongly Coupled Thermal Dynamics,' where part of this work was completed.
This publication has emanated from research conducted with the financial support of Taighde Éireann – Research Ireland under Grant number SFI-22/FFP-P/11444.

\appendix

\section{Coefficients of the Recurrence Relation}\label{app:coeff-table}

In this appendix, we tabulate the explicit expressions for the coefficients \( C_{a,b,c}^{i,j,k} \), indexed by \( a, b, c \in \{0,1,2\} \), and expressed as functions of \( i, j, k \), and the scaling dimension \( \Delta_\phi \). These coefficients appear in the expansion of the operator \( \widetilde{D} \) acting on the basis functions \( f_{i,j,k} \), as defined in the main text in \eqref{Df}.

\[
\begin{array}{c|c}
\textbf{(a, b, c)} & \textbf{\( C_{a,b,c}^{i,j,k} \)} \\
\hline
(0, 0, 0) & 0 \\
(0, 0, 1) & 0 \\
(0, 0, 2) & 2 j (1 + 2 j) \\
(0, 1, 0) & 0 \\
(0, 1, 1) & 4 (4 i - k)(k - \Delta_\phi) \\
(0, 1, 2) & 4 (2 i - j - k)(-2 + 2 i - j - k + \Delta_\phi) \\
(0, 2, 0) & 0 \\
(0, 2, 1) & 0 \\
(0, 2, 2) & 0 \\
\hline
(1, 0, 0) & 0 \\
(1, 0, 1) & 0 \\
(1, 0, 2) & -2 j (1 + 2 j) \pi^4 \\
(1, 1, 0) & -8 \pi^4 (-1 + k - \Delta_\phi)(k - \Delta_\phi) \\
(1, 1, 1) & -2 \pi^4 (k - \Delta_\phi)(3 + 16 i - 4 j - 8 k + 4 \Delta_\phi) \\
(1, 1, 2) & -\pi^4 \left(32 i^2 + 8(j + k)(1 + j + k) - 16 i (1 + 2 j + 2 k - \Delta_\phi) - 8 (j + k)\Delta_\phi + \Delta_\phi^2 \right) \\
(1, 2, 0) & -4 \pi^4 (-1 + k - \Delta_\phi)(k - \Delta_\phi) \\
(1, 2, 1) & 0 \\
(1, 2, 2) & 0 \\
\hline
(2, 0, 0) & 0 \\
(2, 0, 1) & 0 \\
(2, 0, 2) & 0 \\
(2, 1, 0) & 4 \pi^8 (-1 + k - \Delta_\phi)(k - \Delta_\phi) \\
(2, 1, 1) & -4 \pi^8 (k - \Delta_\phi)(-1 - 4 i + 2 j + 2 k - \Delta_\phi) \\
(2, 1, 2) & \pi^8 (4 i - 2(j + k) + \Delta_\phi)^2 \\
(2, 2, 0) & 0 \\
(2, 2, 1) & 0 \\
(2, 2, 2) & 0 \\
\end{array}
\]

\section{Pad\'e-Borel vs. PDE: Double Trace Coefficients Revisited}\label{app:double-trace coefficients matching}

In this appendix, we describe an alternative method for determining the double-trace thermal coefficients that appear in the thermal two-point function to the one presented in Section \ref{SS:Matching}. In the main text we estimate the thermal coefficients of higher double-trace operators one by one by taking derivatives with respect to $\tau$, so that the dependence of the two-point function on the previous coefficients drops out. Another possibility would be to estimate several coefficients at the same time by making a multi-variable fit of the numerical solution $g_\beta(\tau)$. Here we shall follow a slightly different approach. In the first step, the double-trace coefficient $a_{[\phi\phi]_{2\Delta_\phi}}$ is estimated as in Section \ref{SS:Matching}, which gives the result \eqref{leading-dt-coefficient}. Next, we aim to estimate the second coefficient $a_{[\phi\phi]_{2\Delta_\phi+2}}$. In order to do this, we write the truncated conformal block decomposition of $g_\beta(\tau)$, treating $a_{[\phi\phi]_{2\Delta_\phi+2}}$ as unknown. For the other thermal coefficients we use the values obtained in Sections \ref{SS:stress-tensor-recursion} and \ref{SS:Padé-Borel}. Finally, for $a_{[\phi\phi]_{2\Delta_\phi}}$, we use the estimate \eqref{leading-dt-coefficient}. That is, we write
\begin{align}
    \nonumber
 &\frac{1}{\tau^2}\bigg(g_\beta(\tau) - \frac{1}{\tau^3} - \frac{1.107}{\beta^3} -\frac{3\tau}{80}\frac{\pi^4}{\beta^4}  - \frac{38.28}{\beta^7}\tau^4 + \frac{479 \tau^5}{268800}\frac{\pi^8}{\beta^8} - \frac{72.76}{\beta^9} \tau^6 - \\
 \nonumber &\frac{56.86}{\beta^{11}} \tau^8 - \frac{79063 \tau^9}{1107025920}\frac{\pi^{12}}{\beta^{12}} - \frac{66.29}{\beta^{13}}\tau^{10} - \frac{312.6}{\beta^{15}}\tau^{12} + \frac{103234783\tau^{13}}{35670078259200}\frac{\pi^{16}}{\beta^{16}}\bigg) \approx \\ & \frac{a_{[\phi\phi]_{2\Delta_\phi+2}}}{\beta^5} + o(\tau^{12})\ .\label{fit3}
\end{align}
The remainder of the method is similar as in the main text. On the left-hand side of \eqref{fit3} we substitute for $g_\beta(\tau)$ the numerical solution of the PDE. It is observed that the resulting LHS gives an approximately constant function. The coefficient $a_{[\phi\phi]_{2\Delta_\phi+2}}$, which should be the value of this constant, is estimated at the point where the slope of the LHS vanishes. We apply the same method for other coefficients and also for $\Delta_\phi = \frac52$. Our results are shown in Table \ref{table-results-32}. We see that the relative discrepancy between the Pad\'e-Borel and PDE estimates is less than $0.003\%$ for the first two lowest coefficients $a_{[\phi\phi]_{2\Delta_\phi+2}}$, $a_{[\phi\phi]_{2\Delta_\phi+4}}$ and less than $0.1\%$ for the next two coefficients.

\begin{table}
\centering
\begin{tabular}{ | c | c | c | c | c |}
 \hline
  $\Delta_\phi = \frac32$ & Pad\'e-Borel & PDE & GFF & AM  \\ 
 \hline
 $[\phi\phi]_{2\Delta_\phi+2}$ & 7.686 & 7.687 & 12.44 & 7.549 \\
 \hline
 $[\phi\phi]_{2\Delta_\phi+4}$ & 38.28 & 38.29 & 30.25 & 37.60 \\
 \hline
 $[\phi\phi]_{2\Delta_\phi+6}$ & 72.76 & 72.77 & 56.11 & 71.54 \\
 \hline
 $[\phi\phi]_{2\Delta_\phi+8}$ & 56.86 & 56.9 & 90.04 & 58.91 \\
 \hline
\end{tabular}
\caption{Double-trace coefficients using Pad\'e-Borel and numerical solution to PDE.}
\label{table-results-32}
\end{table}

\begin{table}
\centering
\begin{tabular}{ | c | c | c | c | c |}
 \hline
  $\Delta_\phi = \frac52$ & Pad\'e-Borel & PDE & GFF & AM  \\ 
 \hline
 $[\phi\phi]_{2\Delta_\phi+2}$ & 37.67 & 37.67 & 30.25 & -129.7 \\
 \hline
 $[\phi\phi]_{2\Delta_\phi+4}$ & 67.36 & 67.3 & 140.3 & -324.0 \\
 \hline
 $[\phi\phi]_{2\Delta_\phi+6}$ & 851.8 & 851 & 420.2 & 1644 \\
 \hline
 $[\phi\phi]_{2\Delta_\phi+8}$ & 2505 & 2500 & 990.1 & 4487 \\
 \hline
\end{tabular}
\caption{Double-trace coefficients for $\Delta_\phi=5/2$.}
\end{table}

It is seen that the Pad\'e-Borel and PDE estimates obtained using this method are significantly closer to each other than the estimates coming from the method described in Section \ref{SS:Matching}. This is not surprising for the following reason. Let us focus on the coefficient  $a_{[\phi\phi]_{2\Delta_\phi+2}}$. The value reported in the PDE column of Table \ref{table-results-32} is obtained by fitting the left-hand side of \eqref{fit3}. This equation includes the coefficient $ a_{[\phi\phi]_{2\Delta_\phi}}$, whose value was estimated in \eqref{leading-dt-coefficient} by fitting the left-hand side of \eqref{fit}. This equation, in turn, depends on the Pad\'e-Borel estimate of the coefficient $a_{[\phi\phi]_{2\Delta_\phi+2}}$.

In summary, this appendix uses the Pad\'e-Borel estimate of the double-trace coefficient $ a_{[\phi\phi]_{2\Delta_\phi+2}}$ as input to a PDE-based method, which then yields an estimate for the same coefficient. As a result, this procedure can yield an estimate of $a_{[\phi\phi]_{2\Delta_\phi+2}}$ that is biased towards the Pad\'e-Borel value used as an input. In contrast, in Section \ref{SS:Matching}, we do not input the information about the PB estimate in producing the PDE estimate.

\bibliographystyle{JHEP}
\bibliography{bibliography} 

\end{document}